\documentclass[10pt]{article}

\usepackage[T1]{fontenc}
\usepackage[utf8]{inputenc}
\usepackage{lmodern}
\usepackage{microtype}
\usepackage[a4paper,margin=23mm]{geometry}
\usepackage{amsmath,amssymb}
\usepackage{graphicx}
\usepackage{booktabs}
\usepackage{tabularx}
\usepackage{array}
\usepackage{siunitx}
\usepackage{xcolor}
\usepackage{caption}
\usepackage{float}
\usepackage[numbers,sort&compress]{natbib}
\usepackage[hidelinks]{hyperref}
\usepackage[nameinlink,noabbrev]{cleveref}

\graphicspath{{figures/}}
\definecolor{navy}{HTML}{17365D}
\definecolor{teal}{HTML}{167D8D}
\hypersetup{
  colorlinks=true,
  linkcolor=navy,
  citecolor=teal,
  urlcolor=teal,
  pdftitle={Separating Weight and Likelihood Tail Effects in Bayesian Neural Groundwater Inversion},
  pdfauthor={Ke-Ming Shen}
}
\title{\textbf{Separating Weight and Likelihood Tail Effects\\in Bayesian Neural Groundwater Inversion}}
\author{Ke-Ming Shen$^{a,b,*}$\\[0.5em]
\small $^{a}$School of Science, East China University of Technology, Nanchang 330013, China\\
\small $^{b}$Key Laboratory of Quark and Lepton Physics (MOE), Central China Normal University,\\
\small Wuhan 430079, China\\
\vspace{0.2em}
\small $^{*}$Corresponding author: \href{mailto:shen_keming@ecut.edu.cn}{shen\_keming@ecut.edu.cn}\\
\small ORCID: \href{https://orcid.org/0000-0002-0727-3437}{0000-0002-0727-3437}}
\date{}

\begin{document}
\maketitle

\begin{abstract}
Groundwater inversion is commonly built on Gaussian assumptions for both network weights and output errors. The assumptions are convenient, but it is unclear which one should be relaxed when hydraulic conductivity is heavy tailed. Here, Gaussian and q-Gaussian distributions were assigned independently to variational weights and the conditional output likelihood in a $2\times2$ Bayesian neural design. Hydraulic conductivity, porosity, and dispersion were inferred jointly from sparse head and transport observations. Four q regimes ($q=1.0$, 1.2, 1.5, and 1.8) were evaluated with five independently generated data sets and matched training seeds, giving 130 held-out cases per regime and 500 posterior draws per case. At $q=1.5$, q-Gaussian weights with a Gaussian likelihood reduced log-conductivity CRPS by 2.85\% (95\% paired CI 0.82--4.88) and calibration error by 3.72 percentage points. Conductivity RMSE was unchanged, while 95\% interval width increased by 9.86\%. At $q=1.8$, assigning q-Gaussian tails to both components produced no stable accuracy gain and increased interval width by 138.46\%. The Gaussian-generated porosity and dispersion controls showed negligible accuracy changes. Moderate weight-side tails can thus improve distributional prediction, whereas an overly heavy output likelihood trades sharpness for coverage. Separating these two roles provides a clearer basis for uncertainty modeling in groundwater inversion.
\end{abstract}


\section{Introduction}

Hydraulic conductivity controls groundwater flow and strongly affects solute migration. Field measurements rarely resolve its spatial variability at the scale required by a numerical model. Conductivity is therefore inferred from hydraulic heads, concentration histories, or both. The resulting inverse problem is nonlinear and often non-unique, so a single reconstructed field gives an incomplete account of the evidence \citep{carrera2005}. Ensemble filters and smoothers offer practical approximations to Bayesian updating \citep{evensen1994,emerick2013}. More recently, neural models have made high-dimensional inversion and surrogate-based uncertainty propagation less expensive \citep{tartakovsky2020,lykkegaard2021,wang2021,han2022,zhou2022,guo2023pinn,guo2023predictive,wang2023active,zhang2024,chen2025,triplett2025}.

Bayesian neural networks (BNNs) provide one route to predictive uncertainty by placing distributions on network weights \citep{blundell2015}. Gaussian variational weights and a Gaussian conditional likelihood are the usual choices. They are easy to optimize, yet they may be too restrictive when aquifer properties or unresolved inverse-map residuals are heavy tailed. Point accuracy and probability calibration must also be judged separately; an accurate median does not guarantee useful uncertainty bounds \citep{abdar2021,psaros2023}. The q-Gaussian family offers a controlled extension. It reduces to a Gaussian at $q=1$ and develops polynomial tails for $1<q<3$ \citep{tsallis1988}. Related nonextensive formulations have been examined elsewhere \citep{shen2017,kang2023}. In the present work, however, the q-Gaussian is used only as a probability family, with no thermodynamic interpretation assigned to the aquifer.

The term ``heavy-tailed BNN'' hides two distinct choices. Heavy tails may be placed in the variational weight distribution, in the conditional output likelihood, or in both. A broad weight distribution changes the ensemble of inverse maps and mainly affects epistemic uncertainty. A heavy-tailed likelihood changes conditional residuals and acts directly on predictive spread. The two mechanisms are not interchangeable. If they are combined without separate checks, high coverage may simply reflect wide intervals, and a favorable result from one training run may be mistaken for a general improvement.

To separate these effects, Gaussian and q-Gaussian choices were assigned independently to weights and likelihood in a factorial groundwater-inversion experiment. Exact agreement with the Gaussian model was retained at $q=1$. Each non-Gaussian setting was then repeated with five independently generated data sets and matched training seeds. Proper scores, multi-level calibration, interval width, clipping sensitivity, and contaminated-input tests were considered together. The analysis asks whether heavy tails improve point or distributional accuracy, where those tails should be placed, and whether improved coverage is obtained without an unacceptable loss of sharpness.

\section{Methods}

\subsection{q-Gaussian family}

The q-exponential and a location--scale q-Gaussian are
\begin{align}
\exp_q(z) &= \left[1+(1-q)z\right]_+^{1/(1-q)},\\
p_q(x\mid\mu,s) &= Z_q(s)^{-1}\exp_q\left[-\frac{(x-\mu)^2}{2s^2}\right].
\end{align}
The Gaussian is recovered continuously at $q=1$. For $1<q<3$, the normalized distribution is equivalent to a Student-t law with degrees of freedom
\begin{equation}
\nu=\frac{3-q}{q-1}
\end{equation}
and Student-t scale $s\sqrt{2/(3-q)}$. This representation was used for exact sampling and log-density evaluation. Weight-noise draws were clipped to $[-12,12]$ in the primary experiment to stabilize stochastic optimization. Thresholds of 6 and 24 were evaluated separately. The mathematical variance is undefined for $q\ge5/3$; medians, quantile intervals, and continuous ranked probability scores (CRPS) were consequently preferred to variance-dependent summaries.

\subsection{Synthetic flow and transport system}

A dimensionless domain was discretized on a $12\times12$ grid. Spatially correlated $\log(K)$ fields were generated from q-Gaussian innovations with scale 0.8 and a two-cell correlation length. Porosity $\phi$ and log-dispersion $\log(D)$ were generated from correlated Gaussian innovations, providing negative controls for unnecessary tail enrichment. The means were $\phi=0.30$ and $D=0.012$; porosity was clipped to $[0.08,0.55]$.

Steady hydraulic head satisfied
\begin{equation}
\nabla\cdot(K\nabla h)=0,
\end{equation}
with fixed heads of 1 and 0 at the left and right boundaries and no-flow conditions at the remaining boundaries. Darcy pore velocity was $\mathbf{v}=-K\nabla h/\phi$. A continuous inlet plume was propagated with
\begin{equation}
\frac{\partial C}{\partial t}=\nabla\cdot(D\nabla C)-\mathbf{v}\cdot\nabla C.
\end{equation}
Twelve interior sensors supplied head and concentrations at dimensionless times 0.05, 0.10, and 0.20; three outlet breakthrough values were added. The input contained 51 observations, and the target contained $\log(K)$, $\phi$, and $\log(D)$ at 144 cells each.

For each $q\in\{1.0,1.2,1.5,1.8\}$, five independent data sets of 256 cases were generated with seeds 13, 29, 42, 71, and 101. Within each seed, the same data and split were used for every model variant. Each data set was divided into 204 training, 26 validation, and 26 test cases.

\subsection{Factorial Bayesian neural model}

The standardized observation vector was mapped through two 128-unit SiLU layers. Each Bayesian linear layer returned sampled weights, and the output was split into a conditional location and a positive scale for all 432 targets. The objective was
\begin{equation}
\mathcal{L}=-\mathbb{E}_{q_w(w)}[\log p_{q_\ell}(\mathbf{y}\mid\mathbf{x},w)]
+\beta\,\mathrm{KL}[q_w(w)\Vert p_{q_w}(w)]/N_{\mathrm{train}},
\end{equation}
where $q_w$ controls weight tails and $q_\ell$ controls the conditional output likelihood. The latter describes residual uncertainty in the learned inverse mapping; it is not a physical measurement-error likelihood.

Four variants were compared within every non-Gaussian regime: Gaussian weights and likelihood (G--G), q-Gaussian weights with Gaussian likelihood (q--G), Gaussian weights with q-Gaussian likelihood (G--q), and q-Gaussian weights and likelihood (q--q). At $q=1$, all definitions reduce to G--G. Adam was used with learning rate $10^{-3}$, batch size 32, weight decay $10^{-6}$, gradient clipping at 5, and a maximum of 250 epochs. The KL factor was annealed over 30 epochs, and training stopped after 35 epochs without validation improvement. All paired variants shared the same initialization and minibatch seed.

\subsection{Evaluation and paired inference}

Five hundred predictive samples were generated for every held-out case. Posterior medians were used for point metrics. Hydraulic-conductivity accuracy was assessed by RMSE and global structural similarity. Distributional accuracy was assessed in latent space using empirical CRPS \citep{gneiting2007},
\begin{equation}
\mathrm{CRPS}(F,y)=\mathbb{E}|X-y|-\frac{1}{2}\mathbb{E}|X-X'|.
\end{equation}
Central interval coverage was calculated at nominal levels 0.50, 0.80, 0.90, and 0.95. Calibration error was the mean absolute difference between empirical and nominal coverage across these levels; 95\% interval width quantified sharpness. Metrics were first averaged over the 26 test cases within a seed and then summarized over five seeds. Relative changes were paired by seed. Exploratory 95\% confidence intervals used a Student-t interval with four degrees of freedom.

At $q=1.8$, the q--q model was retrained with weight-noise clipping thresholds of 6 and 24, in addition to the primary threshold of 12. A separate test-time stress experiment contaminated 0\%, 5\%, or 10\% of observation entries with three-scale Student-t noise ($\nu=3$). Because the inverse model was trained on clean inputs, this experiment is interpreted as an external robustness diagnostic, not as a likelihood-identification test. Full tables and stress results are provided in Online Resource 1.

OpenAI Codex assisted with code refactoring, repeated-experiment orchestration, and language editing. All model runs, statistical summaries, figures, and manuscript statements were checked by the author, who takes responsibility for the final work.

\section{Results}

\subsection{Numerical and experimental verification}

Numerical integration confirmed normalization of the implemented q-Gaussian over a broad finite domain. At $q=1$, sampled moments agreed with a standard normal distribution, and identical outputs were obtained from the q-Gaussian and Gaussian code paths. The maximum residual of the first discrete groundwater-flow solution was $5.38\times10^{-7}$. All 12 automated tests, including calibration and interval-width checks, passed. Sixty-five primary models were fitted: five G--G models at $q=1$ and four variants for each of three non-Gaussian regimes and five seeds. A further ten models supported clipping sensitivity analysis. No non-finite metric was observed.

\subsection{Weight and likelihood effects}

\begin{table}[H]
\centering
\caption{Five-seed hydraulic-conductivity results. Values are mean (standard deviation). Lower RMSE, CRPS, calibration error, and width are preferred, except that width must be interpreted jointly with coverage.}
\label{tab:ablation}
\scriptsize
\begin{tabular}{llccccc}
\toprule
$q$ & Variant & $K$ RMSE & $\log(K)$ CRPS & Coverage$_{0.95}$ & Width$_{0.95}$ & Calibration error\\
\midrule
1.2 & G--G & 1.323 (1.072) & 0.198 (0.025) & 0.901 & 1.209 & 0.053\\
    & q--G & 1.253 (1.104) & 0.199 (0.028) & 0.912 & 1.255 & 0.040\\
    & G--q & 1.259 (1.079) & 0.204 (0.031) & 0.909 & 1.306 & 0.052\\
    & q--q & 1.242 (1.095) & 0.200 (0.030) & 0.917 & 1.307 & 0.042\\
\addlinespace
1.5 & G--G & 0.981 (0.341) & 0.218 (0.018) & 0.890 & 1.263 & 0.058\\
    & q--G & 0.983 (0.346) & 0.212 (0.020) & 0.927 & 1.387 & 0.020\\
    & G--q & 0.994 (0.352) & 0.220 (0.018) & 0.929 & 1.612 & 0.044\\
    & q--q & 0.996 (0.352) & 0.213 (0.021) & 0.957 & 1.734 & 0.014\\
\addlinespace
1.8 & G--G & 0.775 (0.142) & 0.192 (0.006) & 0.930 & 1.334 & 0.028\\
    & q--G & 0.757 (0.167) & 0.185 (0.014) & 0.979 & 1.643 & 0.082\\
    & G--q & 0.822 (0.126) & 0.205 (0.009) & 0.993 & 2.869 & 0.045\\
    & q--q & 0.768 (0.169) & 0.193 (0.015) & 0.999 & 3.144 & 0.097\\
\bottomrule
\end{tabular}
\end{table}

At $q=1.2$, point-metric changes varied widely among data sets, and CRPS did not improve. A clearer pattern appeared at $q=1.5$. Relative to G--G, q--G changed $K$ RMSE by only $-0.06\%$ (95\% CI $-1.52$ to 1.40) but reduced $\log(K)$ CRPS by 2.85\% (0.82 to 4.88). Calibration error fell by 3.72 percentage points (1.83 to 5.62), while interval width increased by 9.86\% (6.46 to 13.26). Moderate weight-side tails improved the predictive distribution without materially changing the posterior median.

The strong single-seed accuracy improvement previously suggested at $q=1.8$ was not stable under replication. For q--q, the mean RMSE reduction was 1.19\% with a 95\% CI from $-10.37$ to 12.75, and the mean CRPS change was unfavorable ($-0.88\%$, CI $-10.16$ to 8.40). Its 95\% interval width increased by 138.46\% (100.15 to 176.77), and calibration error worsened by 6.87 percentage points. The q--G variant retained the lowest mean RMSE and CRPS at this q, but its paired intervals also crossed zero and its high coverage was obtained with a 24.69\% wider interval.

\begin{figure}[H]
  \centering
  \includegraphics[width=\textwidth]{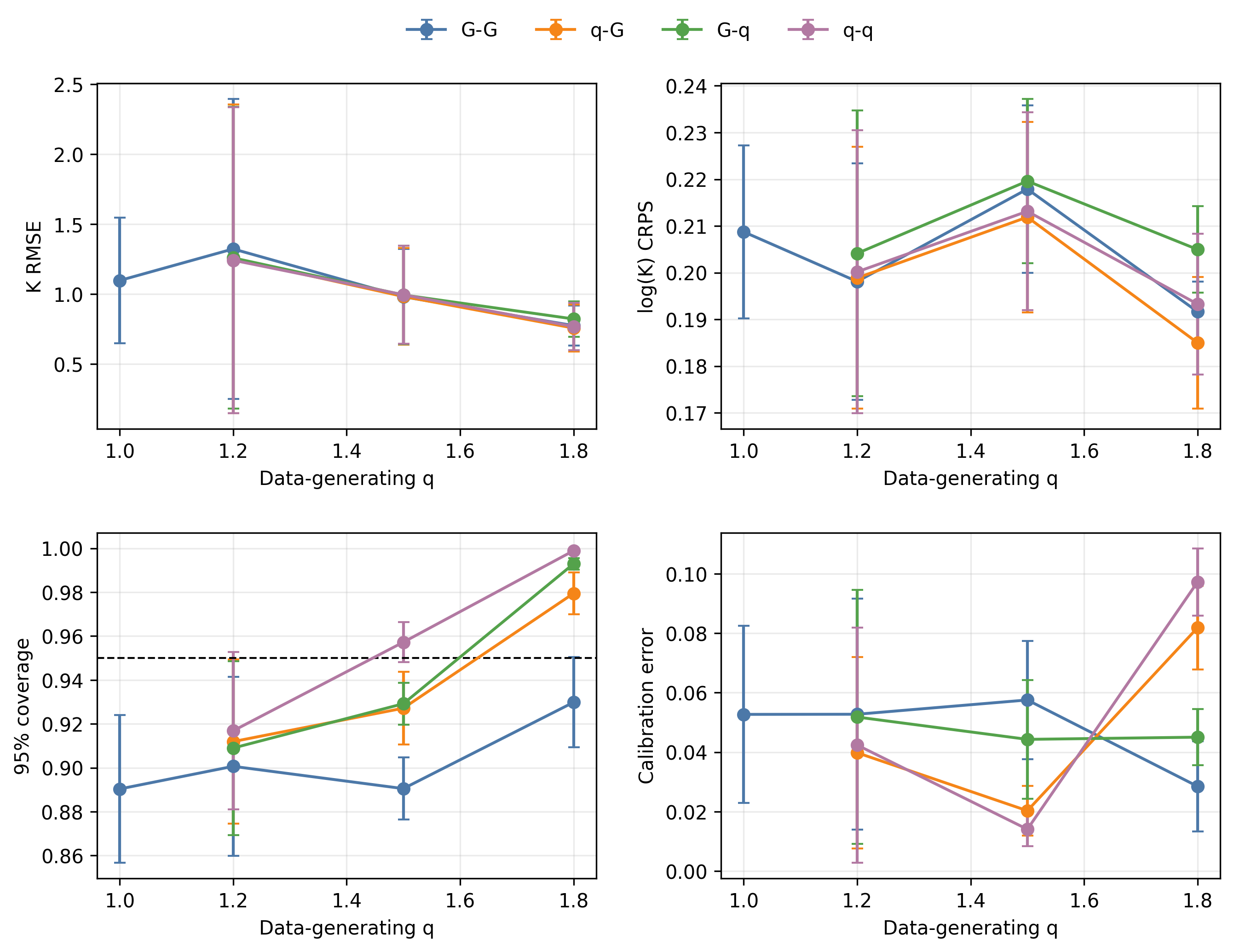}
  \caption{Five-seed means and standard deviations for the factorial variants. Comparisons are made within each data-generating q; connecting lines only guide the eye because independently generated regimes need not have monotonic difficulty.}
  \label{fig:ablation}
\end{figure}

\subsection{Calibration, sharpness, and negative controls}

\begin{figure}[H]
  \centering
  \includegraphics[width=0.70\textwidth]{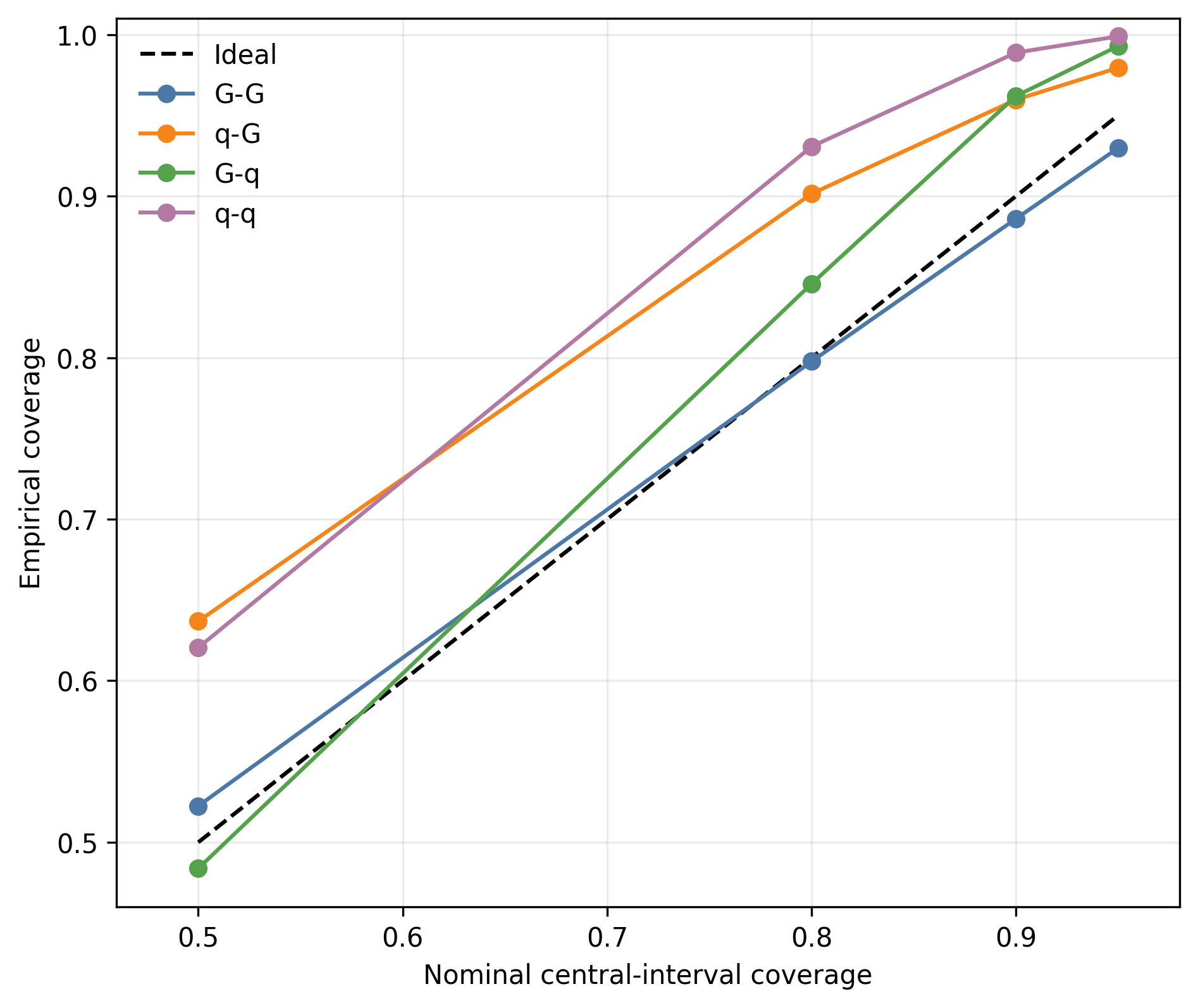}
  \caption{Mean empirical central-interval coverage at $q=1.8$. Curves above the diagonal are over-dispersed.}
  \label{fig:calibration}
\end{figure}

The calibration curve at $q=1.8$ exposes the source of nominally high coverage (\cref{fig:calibration}). G--G remained close to the diagonal. G--q and q--q were above it at every level and approached complete coverage at the 95\% level. Their intervals were respectively 115\% and 138\% wider than G--G in the paired analysis. Thus, coverage alone would have incorrectly favored the heaviest predictive likelihood.

Porosity and dispersion were generated without heavy-tailed innovations and served as negative controls. Across variants, their five-seed RMSE and CRPS changed little. At $q=1.8$, for example, porosity RMSE was 0.03629 for G--G and 0.03605 for q--G; dispersion RMSE was 0.003450 and 0.003440. Adding a q likelihood raised 95\% coverage for both parameters to about 0.99 but did not reduce point error. A more flexible uncertainty family did not compensate for limited information about the jointly inferred properties.

\subsection{Sensitivity and contaminated-observation stress}

For q--q at $q=1.8$, clipping thresholds of 6, 12, and 24 produced mean $K$ RMSE values of 0.796, 0.768, and 0.795 and latent CRPS values of 0.197, 0.193, and 0.201. All three settings remained strongly over-dispersed: coverage exceeded 0.998, and interval widths ranged from 3.13 to 3.45. The result was not tied to the primary clipping threshold.

Under 5\% contaminated input entries, q--q reduced mean RMSE from 1.021 to 0.900 and latent CRPS from 0.268 to 0.252 relative to G--G. At 10\% contamination, q--q retained lower CRPS (0.333 versus 0.377) but its physical-space RMSE became less stable across seeds. These results indicate some probabilistic robustness but do not establish an observation-error mechanism, because the q likelihood is applied to the inverse-model output. Median and interquartile summaries, full tables, and figures are reported in Online Resource 1.

\section{Discussion}

\subsection{Tail placement is a modeling decision}

The factorial results show that the q extension cannot be treated as one indivisible model choice. Making every stochastic component heavy tailed did not produce the most reliable gain. At $q=1.5$, weight-side tails improved CRPS and calibration with a modest loss of sharpness; likelihood-side tails widened intervals much more. Weight tails change the sampled inverse functions and can accommodate parameter heterogeneity without forcing every conditional residual to be extreme. Likelihood tails act directly on predictive draws. At $q=1.8$, their infinite-variance form produced intervals so conservative that useful differences among fields were partly obscured.

A single run gave a different impression. The original seed at $q=1.8$ suggested a large RMSE gain, whereas five independent data and training seeds reduced the mean effect to about one percent. Its confidence interval included both benefit and harm. Seed-level pairing therefore changed the scientific conclusion rather than merely adding detail.

\subsection{Calibration must be coupled to sharpness}

Coverage is meaningful only when read alongside interval width and calibration at more than one nominal level. The q--q model attained nearly complete 95\% coverage at $q=1.8$, but its intervals were more than twice as wide as those of G--G. Proper scoring did not improve either. Such a posterior makes many fields plausible while offering little discrimination. For heavy-tailed groundwater inversion, CRPS, interval width, and calibration curves should be reported together.

\subsection{Identifiability limits distributional flexibility}

The behavior of $\phi$ and $D$ also clarifies the role of the data. Sparse heads and three early concentration snapshots constrain some conductivity patterns, yet compensation remains possible among conductivity, porosity, and dispersion. Flexible tails changed interval coverage but did not add information. Better recovery of these properties will require denser temporal sampling, adaptive sensor placement, cross-property geological priors, or lower-dimensional representations. A heavier probability family alone cannot resolve this limitation.

\subsection{Limitations and future work}

The experiment is a controlled synthetic study on a $12\times12$ dimensionless grid. Five seeds capture part of the algorithmic and data-generation variability, but they do not establish transfer to field conditions. The direct inverse network also attaches the conditional likelihood to parameter targets rather than to a physical observation model. A hierarchical extension could estimate separate $q_w$ and $q_\ell$ values under a proper validation score. Tempered heavy-tailed families with finite moments are another option.

The next step is a higher-resolution MODFLOW 6/GWT or equivalent benchmark with dimensional units, conservation checks, heterogeneous boundary conditions, and an independent geological configuration. Physics-matched ensemble smoothers and neural models should use the same forward simulator, or a nonlinear surrogate validated against it. Observation-error contamination should be represented in the physical likelihood. Sensor locations and sampling times can then be optimized with the posterior, linking improved uncertainty estimates to practical reductions in ambiguity.

\section{Conclusions}

Gaussian and q-Gaussian laws were assigned independently to variational weights and the conditional output likelihood. This factorial design separated two roles that are often grouped under the label of a heavy-tailed BNN. The Gaussian limit was preserved exactly, and each non-Gaussian regime was repeated on five independently generated data sets. Replication removed the apparent large gain seen in one seed and replaced it with a more stable, mechanism-based result.

At $q=1.5$, moderate q-Gaussian weight tails improved log-conductivity CRPS and multi-level calibration without changing point RMSE. The accompanying increase in interval width was about 10\%. Heavier tails were not uniformly helpful. At $q=1.8$, assigning q-Gaussian tails to both weights and likelihood produced no stable accuracy gain and more than doubled interval width. The Gaussian-generated porosity and dispersion controls showed negligible accuracy changes. Tail enrichment was useful only when it matched a real distributional mismatch.

The main contribution is a reproducible way to decide where heavy tails belong. Weight-side and likelihood-side effects are evaluated separately, and apparent gains caused by seed variability or overly wide intervals can be identified. This directly answers the question posed at the outset: tail placement matters as much as tail weight, and coverage alone is not enough. Future work will extend the comparison to adaptive q estimation, finite-moment robust families, physics-based benchmarks, and observation designs that balance coverage with posterior sharpness.

\section*{Acknowledgements}
The author thanks Si-Qi Yin, Yang Liu, Yi-Xuan Liu, Chen-Yue Yu, and colleagues for helpful discussions that informed the development and presentation of this work.

\bibliographystyle{unsrtnat}
\bibliography{references}

\section*{Statements and Declarations}

\subsection*{Funding}
This work is supported in part by Innovative Platform for Quark Matter Science and Advanced Detection Technology funded under the Ministry of Education's ``Two-New'' Initiative (No. QLPL2026P01).

\subsection*{Competing interests}
The author has no relevant financial or non-financial interests to disclose.

\subsection*{Author contributions}
Ke-Ming Shen: conceptualization, methodology, software, formal analysis, investigation, visualization, writing--original draft, and writing--review and editing. The author read and approved the final manuscript.

\subsection*{Data availability}
All synthetic data, configuration files, result tables, and figures supporting this study are included in the accompanying reproducibility archive and Online Resource 2. No confidential or third-party data were used. A public archival DOI may be added before or after acceptance.

\subsection*{Code availability}
The source code used to generate the synthetic systems, train all variants, calculate metrics, and reproduce the figures is included in Online Resource 2. Fixed seeds and machine-readable configuration files are provided.

\subsection*{Ethics approval}
Not applicable. The study used synthetic numerical data and involved no human participants or animals.

\end{document}